\documentclass[prl,aps,epsfig,twocolumn,showpacs]{revtex4}
\usepackage{amsmath}

\input{tcilatex}

\begin{document}

\title{Propagation of Fronts and Information in Dispersive Media }
\author{Shi-Yao Zhu$^{1,2}$, Ya-Ping Yang$^{1,2}$, Li-Gang Wang$^{2}$,
Nian-Hua Liu$^{2,3}$, and M. Suhail Zubairy$^{4}$}
\affiliation{$^{1}$Department of Physics, Tongji University, Shanghai 200092, China}
\affiliation{$^{2}$Department of Physics, Hong Kong Baptist University, Kowloon Tong,
Hong Kong }
\affiliation{$^{3}$Department of Physics, Nanchang University, Nanchang 330047, China}
\affiliation{$^{4}$Institute for Quantum Studies and Department of Physics, Texas A\&M
University, College Station, TX 77843}
\date{\today }

\begin{abstract}
We present a general proof based on Kramers-Kronig relations that, in a
normal or anomalous dispersive linear medium, any (discontinuitynonanalytic
disturbance) in an electromagnetic pulse can not propagate faster than the
phase velocity, $c$. Consequently the information carried by the
discontinuity (nonanalytical disturbance) can not be transmitted
superluminally.
\end{abstract}

\preprint{APS/123-QED}
\maketitle


According to the Einstein's theory of special relativity, the speed of any
moving object can not exceed the speed of light in vacuum $c$. However it is
well-known that the group velocity of a light pulse $v_{g}$ can exceed $c$
in an anamolously dispersive medium \cite{RYChiao1}. This interesting effect
is a result of interference of different frequency components of the light
pulse \cite{1Dogariu,2Dogariu}. The superluminal phenomenon disappears when
the pulse loses the coherence \cite{WangLG}.

Sommerfeld and Brillouin observed in 1914 that a superluminal group velocity
does not violate causality \cite{som,bril,bril2}. They observed that the
front velocity, (the velocity of a sharp nonanalytic discontinuity in a
light pulse), should be used as the signal velocity at which information is
transmitted and this velocity does not exceed $c$. The analysis of
Sommerfeld and Brillouin is based on the propagation of an electromagnetic
wave of the form 
\begin{equation}
f(t)=\left\{ 
\begin{matrix}
0 & (t<0) \\ 
\sin (2\pi t/\tau ) & (t>0)%
\end{matrix}
\right.
\end{equation}
through an anamolously dispersive medium with strong absorption,
characterzed by a Lorentzian susceptibilty that is proportional to $\omega
^{-2}$. The sharp begining of the light wave (1) corresponds to the signal.
The debate concerning the information velocity in media still remains due to
progress in experiments. For example, a recent experiment by Wang et al.
reports superluminal propagation in a gain medium with susceptibility
proportional to $\omega ^{-1}$ \cite{LJWang}. It is therefore a matter of
great interest to give a general proof about the causal nature of the
propagation of classical information (carried by the front or
discontinuities) with a velocity less than or equal to $c$.

We note that the \emph{peak superluminal propagation} of a light pulse in a
dielectric medium, when the spectrum of the incident pulse is in the
anomalous dispersion frequency range of the medium, can be derived from
Maxwell's electromagnetic theory. The superluminal propagation is therefore
a classical phenomenon and the question whether the information can be
transmitted with a velocity faster than $c$ needs to be addressed
classically. Any practical pulse must have a beginning (a starting point of
a non-equilibrium process) \cite%
{RYChiao1,Mojahedi,Garrison,Mitchell,Solli,Japha}. Furthermore, any
nonanalytical disturbance (a discontinuity of the field or its first or
higher order derivatives) \cite{Japha,Mojahedi,Mitchell} carries classical
information. The front of the pulse is one of such discontinuities.

In this letter, we prove, based only on Kramers-Kronig relations and the
Maxwell equations, that any nonanalytical disturbance (any discountinuity
including the front) in pulses propagates at the phase velocity in a linear
medium. It should be emphsized that our general and rigorous proof has no
requirement on the form of incident field and the particularity of the
medium.

We consider the propagation of an electromagnetic pulse through a medium
occupying the spapce from $z=0$ to an arbitrary $z>0$ whose response to the
electric field of the light pulse is linear. After passing through the
medium, the light field at the position $z$ and at time $t$ can be written
as 
\begin{equation}
E_{m}\left( z,t\right) =\int_{-\infty }^{+\infty
}dt_{1}E_{0}(0,t_{1})G(t_{1}-t+z/c;z),
\end{equation}
where 
\begin{equation}
G(\xi ;z)=\frac{1}{2\pi }\int_{-\infty }^{+\infty }d\omega e^{i\omega \xi
}e^{i\omega \sigma (\omega )z/c}  \label{F1}
\end{equation}
has the property of the retarded Green function. Here $E_{0}(0,t_{1})$ is
the input field and $\sigma (\omega )=n(\omega )-1+i\kappa (\omega )$ with $%
n(\omega )$ and $\kappa (\omega )$ being the real and imaginary parts of the
complex refractive index, respectively. Since $n(\omega )$ and $\kappa
(\omega )$ satisfy the Kramers-Kronig relations \cite{Loudon1979}, it can be
shown that $\sigma (\omega )$ is an analytic function in the upper half
plane and consequently $\sigma (\omega )$ can be expanded into the form: $%
\sigma (\omega )=\sum\limits_{n=1}^{\infty }A_{n}/\omega ^{n}$ for $\left|
\omega \right| \rightarrow \infty $ , where $A_{n}$ are the expansion
coefficients.

For the pulse propagation through vacuum, $\sigma (\omega )=0$. Therefore
the output field at $z$ is $E_{v}(z,t)=E_{0}(0,t-z/c)$,i.e., the light pulse
propagates with velocity $c$.

For the pulse propagation through a medium, the function $G(\xi ;z)$ as
given by Eq. (3) can be rewritten as 
\begin{eqnarray}
G(\xi ;z) &=&\frac{1}{2\pi }\int_{-\infty }^{+\infty }d\omega e^{i\omega \xi
}\left\{ \left[ e^{i\omega \sigma (\omega )z/c}-e^{i\frac{zA_{1}}{c}}\right]
+e^{i\frac{zA_{1}}{c}}\right\}  \notag \\
&=&e^{i\frac{zA_{1}}{c}}\delta (\xi )+e^{i\frac{zA_{1}}{c}}J(\xi ;z)
\label{F3}
\end{eqnarray}
where 
\begin{equation}
J(\xi ;z)=\frac{1}{2\pi }\int_{-\infty }^{+\infty }d\omega e^{i\omega \xi }%
\left[ e^{-i\frac{zA_{1}}{c}}e^{i\omega \sigma (\omega )z/c}-1\right] .
\label{JJ1}
\end{equation}
We then have 
\begin{equation}
E_{m}\left( z,t\right) =e^{i\frac{zA_{1}}{c}}E_{0}(z,t-z/c)+e^{i\frac{zA_{1}%
}{c}}\bar{E}_{m}(z,t),  \label{EMnew}
\end{equation}
where 
\begin{equation}
\bar{E}_{m}(z,t)=\int_{-\infty }^{t-z/c}dt_{1}E_{0}(0,t_{1})J(t_{1}-t+z/c;z)
\end{equation}
From Eq. (\ref{EMnew}), we see that the output field has two parts: the
first part is an instant response which leads to a time delay $(z/c)$ for
the discontinuities in the field, and the second part is the retarded
response from the medium. In the following, we will prove that the second
term is a continuous function. The discontinuity in the field is therefore
determined by the first term, thus proving that the discontinuity always
prapagates with the phase velocity.

As the integral function in Eq. (\ref{JJ1}) is analytic in the upper half
plane, $J(\xi ;z)$ is equal to $0$ for $\xi >0$ since the integral function $%
\left[ e^{-i\frac{zA_{1}}{c}}e^{i\omega \sigma (\omega )z/c}-1\right]%
\rightarrow 0$ when $\left\vert \omega \right\vert \rightarrow \infty $.

When $\xi =0$, we have 
\begin{eqnarray}
J(\xi ;z) &=&\frac{1}{2\pi }\int_{-\infty }^{+\infty }d\omega \left[ e^{-i%
\frac{zA_{1}}{c}}e^{i\omega \sigma (\omega )z/c}-1\right]  \notag \\
&=&\frac{1}{2\pi }\lim\limits_{R\rightarrow \infty }\int_{C}d\omega \left[
e^{-i\frac{zA_{1}}{c}}e^{i\omega \sigma (\omega )z/c}-1\right]  \notag \\
&=&\frac{1}{2\pi }\lim\limits_{R\rightarrow \infty }\int_{C}d\omega
\sum\limits_{n=1}^{\infty }\frac{B_{n}}{\omega ^{n}}  \notag \\
&=&-\frac{i}{2}B_{1},  \label{JF2}
\end{eqnarray}
where the integrations in the second and third lines are along the open
semicircle $C$ under the condition $R\rightarrow \infty $ (see Fig. 1(a)).
From the first to second lines, we have used that the intagration along the
closed path composed of the real axis and the semicircle $C$ (with $%
R\rightarrow \infty $)\ is zero, because\ the integral, $exp[-iz/c+i\frac{%
\omega \sigma (\omega )z}{c}]-1$, is analytical. \ In the third line we
expanded $exp[-iz/c+i\frac{\omega \sigma (\omega )z}{c}]-1$ for $\left|
\omega \right| \rightarrow \infty $, into $\sum\limits_{n=1}^{\infty
}(B_{n}/\omega ^{n})$, where the coefficients $\{B_{n}\}$ are related to the
coefficients $\{A_{n+1}\}$.

\FRAME{ftFU}{5.8101cm}{7.5959cm}{0pt}{\Qcb{(a) The integral path of Eq. (%
\protect\ref{JF2}); (b) The integral path of Eq. (\protect\ref{JF3}).}}{\Qlb{%
FIGURE1}}{fig1.eps}{\special{language "Scientific Word";type
"GRAPHIC";maintain-aspect-ratio TRUE;display "USEDEF";valid_file "F";width
5.8101cm;height 7.5959cm;depth 0pt;original-width 7.7574in;original-height
11.0627in;cropleft "0.2063";croptop "0.9272";cropright "0.7420";cropbottom
"0.4342";filename 'fig1.eps';file-properties "XNPEU";}}

For $\xi <0$, as shown in Fig. 1(b), we have 
\begin{eqnarray}
J(\xi ;z) &=&\frac{1}{2\pi }\int_{-\infty }^{+\infty }d\omega e^{i\omega \xi
}\left[ e^{-i\frac{zA_{1}}{c}}e^{i\omega \sigma (\omega )z/c}-1\right] 
\notag \\
&=&\frac{1}{2\pi }\oint_{C^{\prime}}d\omega e^{i\omega \xi }\left[ e^{-i%
\frac{zA_{1}}{c}}e^{i\omega \sigma (\omega )z/c}-1\right]  \notag \\
&=&-\frac{1}{2\pi }\left[ \oint_{I}+\oint_{II}+\cdots \right]  \notag \\
&&\left[ e^{-i\frac{zA_{1}}{c}}e^{i\omega \sigma (\omega )z/c}-1\right]
e^{i\omega \xi }d\omega  \notag \\
&=&-\frac{1}{2\pi }\oint_{r}\left[ e^{-i\frac{zA_{1}}{c}}e^{i\omega \sigma
(\omega )z/c}-1\right] e^{i\omega \xi }d\omega  \notag \\
&=&-\frac{1}{2\pi }\oint_{r}\left[ e^{-i\frac{zA_{1}}{c}}e^{i\omega \sigma
(\omega )z/c}-1\right]  \notag \\
&&\times e^{i\omega \xi }re^{i\varphi }id\varphi,  \label{JF3}
\end{eqnarray}
where the step leading to second line is because the function $\left[ e^{-i%
\frac{zA_{1}}{c}}e^{i\omega \sigma (\omega )z/c}-1\right]$ tends to zero
when $\left\vert \omega\right\vert\rightarrow \infty$; in the second line,
the integral is along the closed path $C^{\prime}$ (see Fig. 1(b)); in the
third line the integrals are along the finite numbers of neighborhoods of
the isolated singular points and tangent lines, and these intergrals are
finite; In the fourth line we describe a dashed circle of a radius $r$ which
embraces all these finite sigularities and tangent lines as shown in Fig.
1(b); and in the fifth line, we let $\omega =r\times e^{i\varphi }$.
Therefore, we have the inequality 
\begin{equation*}
\left\vert J(\xi ;z)\right\vert \leq \frac{1}{2\pi }\int_{0}^{2\pi }M\times
rd\varphi =r\times M
\end{equation*}
where $M$ is the maximum value of the function $\left\vert \left[ e^{-i\frac{%
zA_{1}}{c}}e^{i\omega \sigma (\omega )z/c}-1\right] e^{i\omega \xi
}e^{i\varphi }\right\vert $ on the dashed circle of radius $r$. This means
that $J(\xi ;z)$ is bounded. Because the function $J(\xi ,z)$ is bounded,
the second term in Eq. (\ref{EMnew}) is continuous. A discontinuity in the
input field amplitude at time $t_{s}$, will occur in the output field after
passing through the medium only at the time equal to $t_{s}+z/c$. This is
the arrival time of information carried by the discontinuity after passing
through the medium, which propagates with the phase velocity $c$.

The intensity discontinuity might disappear under some specific cases due to
the second term. The reason is that although $|a|^{2}-|b|^{2}\neq 0$, $%
|a+c|^{2}-|b+c|^{2}$ may be equal to zero for complex numbers $a,b,c$. Once
such a discontinuity disappears, there is no intensity discontinuity at any
time, and the information is lost. It should be pointed out that there is no
requirement for the form of the incident field in the above proof and
particularity of the medium except the Kramers-Kronig relations.

Next we consider the propagation of a nonanalytic disturbance in the
derivatives of the field amplitude. Suppose the $0$th to the $(n-1)$th order
derivatives of the field are continuous functions, while the $n$th
derivative has a discontinuity. How does this kind of nonalytical
disturbance propagate? From Eq. (\ref{EMnew}), we obtain 
\begin{equation}
\frac{\partial ^{n}E_{m}\left( z,t\right) }{\partial t^{n}}=e^{i\frac{zA_{1}%
}{c}}\frac{\partial ^{n}E_{0}(z,t-z/c)}{\partial t^{n}}+e^{i\frac{zA_{1}}{c}}%
\frac{\partial ^{n}\bar{E}_{m}(z,t)}{\partial t^{n}}.
\end{equation}
The second term is 
\begin{eqnarray}
\frac{\partial ^{n}\bar{E}_{m}(z,t)}{\partial t^{n}} &=&\frac{1}{2\pi }%
\int_{-\infty }^{+\infty }dt_{1}E_{0}(0,t_{1})\int_{-\infty }^{+\infty
}d\omega (-i\omega )^{n}  \notag \\
&&\times e^{i\omega \xi }\left[ e^{-i\frac{zA_{1}}{c}}e^{i\omega \sigma
(\omega )z/c}-1\right]
\end{eqnarray}
where the integral $\int_{-\infty }^{+\infty }d\omega (-i\omega
)^{n}e^{i\omega \xi }\left[ e^{-i\frac{zA_{1}}{c}}e^{i\omega \sigma (\omega
)z/c}-1\right] $ is equal to $(-i)^{n}\int_{-\infty }^{+\infty }d\omega
e^{i\omega \xi }\underset{n}{\underbrace{\{\omega \{\cdots \{\omega }}[e^{-i%
\frac{zA_{1}}{c}}e^{i\omega \sigma (\omega )z/c}-1]-B_{1}\}\cdots
\}-B_{n}\}+2\pi \sum\limits_{j=1}^{n}(-i)^{j}B_{j}\frac{\partial
^{(n-j)}\delta (\xi )}{\partial t^{(n-j)}}$. Therefore, we finally obtain 
\begin{gather}
\frac{\partial ^{n}E_{m}\left( z,t\right) }{\partial t^{n}}=e^{i\frac{zA_{1}%
}{c}}\sum\limits_{j=0}^{n}(-i)^{j}B_{j}\frac{\partial ^{(n-j)}E_{0}(z,t-z/c)%
}{\partial t^{(n-j)}}  \notag \\
+e^{i\frac{zA_{1}}{c}}\frac{1}{2\pi }\int_{-\infty }^{+\infty
}dt_{1}E(0,t_{1})J_{n}^{\prime }(t_{1}-t+z/c,z).  \label{Emnth}
\end{gather}
where $J_{n}^{\prime }(\xi ,z)=(-i)^{n}\int_{-\infty }^{+\infty }d\omega
e^{i\omega \xi }\underbrace{\{\omega \{\cdots \{\omega }\{e^{\frac{-izA_{1}}{%
c}}\newline
\times e^{\frac{i\omega \sigma (\omega )z}{c}}-1\}-B_{1}\}\cdots \}-B_{n}\}$%
. Following the same reasoning as discussed above, we can again prove that $%
J_{n}^{^{\prime }}(\xi,z)$ is equal to 0 for $\xi >0$ and is bounded for $%
\xi \leq 0$. Therefore, the second term in Eq. (\ref{Emnth}) is a continuous
function of time. The discontinuity in the $n$th order derivative of the
output field solely comes from the first term in Eq. (\ref{Emnth}) (recall
that the 0th to the $(n-1)$th order derivatives are continuous).

We can conclude that the discontinuity in the $n$th order derivative
propagates with the phase velocity $c$. This discontinuity might also be
washed out under certain cases. Once this discontinuity disappears, there is
no discontinuity at any other time (the information is lost). Any initial
discontinuity in the $n$th order derivative (including the 0-th order) never
leads to any new nonanalytical disturbances during the propagation.
Therefore there is one to one correspondance between the disturbances
(information) in the output field and the input field.

\FRAME{ftbpF}{3.2889in}{2.7873in}{0pt}{\Qcb{The intensities and their first
order derivatives of the output fields after passing through the gain medium
(solid lines) and the same distance of vacuum (dashed lines). The input
field is $E(0,t)=2.0\exp [-t^{2}/(2\protect\sigma _{t}^{2})]$ when $t\leq 0,$
$E(0,t)=2.0-5.0t$ when $0<t<0.2$, and $E(0,t)=\exp [-(t-0.2)^{2}/(2\protect%
\sigma _{t}^{2})]$ when $t\geq 0.2.$ The parameter of the medium is the same
as WKD's experiment\protect\cite{LJWang}.}}{}{newfig2.eps}{\special{language
"Scientific Word";type "GRAPHIC";maintain-aspect-ratio TRUE;display
"USEDEF";valid_file "F";width 3.2889in;height 2.7873in;depth
0pt;original-width 2.4561in;original-height 2.0781in;cropleft "0";croptop
"1";cropright "1";cropbottom "0";filename '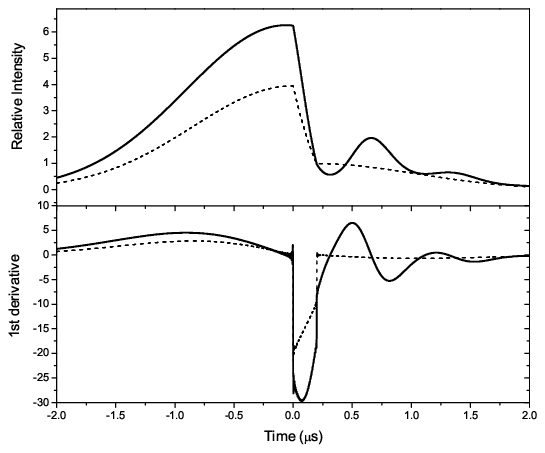';file-properties
"XNPEU";}}

\FRAME{ftbpF}{3.3131in}{2.8184in}{0pt}{\Qcb{The intensities and their 2nd
order derivatives of the output field after passing through the gain medium
(solid lines) and the same distance vacuum (dashed lines). The input field
is $E(0,t)=A[1+\cos(\protect\pi t/2.4)]$ when $-7.2<t<-4.8$ and $4.8<t<7.2$, 
$E(0,t)=A[1.5+0.5\cos(\protect\pi t/2.4)]$ when $-4.8<t<4.8$, and otherwise $%
E(0,t)=0$. The parameter of the medium is the same as WKD's experiment 
\protect\cite{LJWang}.}}{}{newfig3.eps}{\special{language "Scientific
Word";type "GRAPHIC";maintain-aspect-ratio TRUE;display "USEDEF";valid_file
"F";width 3.3131in;height 2.8184in;depth 0pt;original-width
2.4284in;original-height 2.0626in;cropleft "0";croptop "1";cropright
"1";cropbottom "0";filename 'Newfig3.EPS';file-properties "XNPEU";}}

As a numerical example, the propagation time of the discontinuities in the
first order derivative are calculated, see Fig. 2. There are two
nonanalytical disturbances in the first order derivative of the input field
at $t_{1}=0$ and $t_{2}=0.2\mu $s, and other parts of the input field are
analytic. After passing through the gain medium of length $L=6$cm,\cite%
{LJWang} these two discontinuities occur at time $t_{1,2}^{\prime
}=t_{1,2}+L/c$ given by the phase velocity, but not by the group velocity.
Please note that the initial intensity is a parabolic line in the interval $%
(0,0.2)$$\mu$s. The information carried by the discontinuities propagate at
the phase velocity $c$.

In the second example, we calculate the propagation of the nonanalytical
disturbances in the second derivative (see Fig.3). There are two such
discontinities at time $t_{1}=-4.8\mu$s and $t_{2}=4.8\mu$s (see Fig.3). 
The analytical disturbances in the 2nd order derivative of the output field
after passing through the gain medium arrive at $t_{1,2}^{%
\prime}=t_{1,2}+L/c $ (same as passing through vacuum). That is to say,
these two nonanalytical disturbances (the encoded information) propagate
through the gain medium with the phase velocity $c$. From the expression of
the input field, we know that there are two more nonalytical disturbances in
the third derivative of the input intensity at time $t_{3}=-7.2\mu$s and $%
t_{4}=7.2\mu$s. The disturbances in the third derivatives propagate also at
the phase velocity $c $.

In the two examples of numerical calculation, we used a gain medium of the
type considered in the experiment \cite{LJWang} with anomalous dispersion.
Here we would like to note that the Kramers-Kronig relations is applicable
to the gain mediun. To our knowledge there is no dielectric media that do
not obey the Kramers-Kronig relations. We also calculated the propagation
time for the normal dispersive media. The calculations confirm the same
result that the nonanlytical disturbances propagate at the phase velocity,
although the group velocities of these media are smaller than the phase
velocity.

For a practical medium, the background refraction index of the medium always
exists. In this case, we can write $n(\omega )=n_{0}+\bar{n}(\omega )$ ( $%
n_{0}$ is the backgroud refractive index and is larger than unity, and $\bar{%
n}(\omega )\rightarrow 0$ as $|\omega |\rightarrow \infty $). In this case,
we can again show that the propagation of any nonanalytical disturbance is
at the speed of $c/n_{0}$ (the phase velocity).

We have proved that the information carried by the nonanalytical
disturbances in the amplitude or in its any order derivatives (including the
front of a pulse) propagates with the phase velocity, but not the group
velocity. Our proof is based only on Maxwell's electromagnetic theory and
the Kramers-Kronig relations without any condition on whether the medium is
normal or anomalous. We believe this is why Einstein only considered the
phase velocity in his relativity theory.

\acknowledgments The authors gratefully acknowledge the support from RGC and
CRGC from Hong Kong Government, National Science Foundation of China, and
the Air Force Research Laboratories (Rome, New York). 

\end{document}